\documentstyle[preprint,aps]{revtex}
\begin{document}
\draft
\title{{\bf Exact Renormalization Scheme for Quantum Anosov Maps}}
\author{{\bf Itzhack Dana}}
\address{Department of Physics, Bar-Ilan University, Ramat-Gan 52900, Israel}
\maketitle

\begin{abstract}
An exact renormalization scheme is introduced for quantum Anosov maps
(QAMs) on a torus for general boundary conditions (BCs), whose number is
always finite. Given a QAM $\hat{U}$ with $k$ BCs and Planck's constant
$\hbar =2\pi /p$ ($p$ integer), its $n$th renormalization iterate
$\hat{U}^{(n)}={\cal R}^{n}(\hat{U})$ is associated with $k$ BCs for all
$n$ and with a Planck's constant $\hbar ^{(n)}=\hbar /k^{n}$. It is
shown that the quasienergy eigenvalue problem for $\hat{U}^{(n)}$ for
{\em all} $k$ BCs is equivalent to that for $\hat{U}^{(n+1)}$ at some
{\em fixed} BCs, corresponding, for $n>0$, to either strict {\em
periodicity} for $kp$ even or {\em antiperiodicity} for $kp$ odd. The
quantum cat maps are, in general, fixed points of either ${\cal R}$ or
${\cal R}^{2}$. The Hannay-Berry results turn out then to be significant
also for general BCs.\newline
\end{abstract}


Nonintegrable systems whose dynamics can be reduced to a 2D torus in
phase space have attracted much attention in the quantum-chaos
literature. When quantizing such a system on the torus, the admissible
quantum states must satisfy proper boundary conditions (BCs), i.e., they
have to be periodic on the torus up to constant Bloch phase factors
specified by a Bloch wave vector ${\bf w}$. If the Hamiltonian of the
system is periodic in phase space, such as, for example, the kicked
Harper model \cite{l,d1,d2,d3}, its classical dynamics can be reduced to
the toral phase space of one unit cell of periodicity, and all Bloch
wave vectors ${\bf w}$ in some Brillouin zone (BZ) are allowed. When
studying the quantum-chaos problem for such a system, it is natural and
important to consider the sensitivity of the eigenstates to continuous
variation of ${\bf w}$ in the BZ \cite{l,d1,d2,d3}. This sensitivity is
usually strong for eigenstates spread over the chaotic region and weak
for eigenstates localized on stability islands.\newline

In general, however, the Hamiltonian of a system whose dynamics can be
reduced to a torus is {\em not} periodic in phase space. Simple and well
known examples are the purely chaotic, Anosov ``cat maps'' \cite
{va,hb,e,pv,d4,k,ke,cat,n}, whose Hamiltonians are quadratic in the
phase-space variables \cite{ke}. When quantizing these systems, it turns
out that only a {\em finite} set of ${\bf w}$'s in the BZ is allowed
\cite{k,cat,n}, see Eq. (\ref{cw}) below, but this set {\em increases}
with increasing chaotic instability. For a large class of cat maps, the
value ${\bf w=0}$, corresponding to strictly periodic quantum states on
the torus, is allowed. This class of maps was first quantized, for ${\bf
w=0}$, in the well known work of Hannay and Berry \cite{hb}. As a matter
of fact, almost all the investigations of the quantum cat maps have been
confined to this class with ${\bf w=0}$. Recently \cite{n}, the case of
antiperiodic BCs (the quantum state assumes values of opposite signs on
opposite sides of the torus) has been studied in some detail. The
results of Hannay and Berry \cite{hb} and of Keating \cite{ke} revealed
a very atypical feature of quantum cat maps, i.e., the high degeneracy
in their spectra, which increases in the semiclassical limit.\newline

Typical spectral properties, fitting generic eigenvalue statistics, are
already found by quantizing torus maps that are very slight
perturbations of the cat maps \cite{pc,pc1,pc2}. According to Anosov's
theorem \cite{va}, these maps have essentially the same classical
dynamics, in particular they are purely chaotic, as the unperturbed cat
maps, which are structurally stable. This ceases to be the case for
larger perturbations that cause bifurcations generating elliptic islands
\cite{pc2}. However, the quantum BCs for a perturbed cat map are the
same as those of the unperturbed cat map, independently of the size of
the perturbation \cite{kmr}, see Eq. (\ref{cw}) below. Because of this
reason and to simplify terms, general perturbed cat maps are referred to
as Anosov maps in this paper. The importance of these maps is in that
they may be viewed as {\em generic} torus maps on the basis of a general
expression for a smooth torus map derived recently \cite{kmr}, see
below. Understanding the properties of quantum Anosov maps (QAMs) for
general toral BCs is essentially an open problem, since almost all the
investigations have been confined to the strict-periodicity case, ${\bf
w=0}$.\newline

In this paper, we introduce an exact renormalization scheme for QAMs
for general BCs on the torus. We show that the spectrum and eigenstates
of a general QAM for {\em all} BCs, and therefore all its quantum
properties, can be fully reproduced from those of the renormalized QAM
at some special, {\em fixed} BCs. Thus, the general BCs are practically
``eliminated'' by the renormalization. Specifically, consider a QAM
given by the evolution operator $\hat{U}$ quantizing a classical Anosov
map. Quantization on a torus requires a Planck's constant $\hbar$ to
satisfy $2\pi /\hbar =p$, an integer. The finite number of BCs is
denoted by $k$, which depends only on the classical unperturbed cat map.
We define a renormalization transformation ${\cal R}$ generating by
iteration a sequence of QAMs $\hat{U}^{(n)}={\cal R}^{n}(\hat{U})$ on
the same torus. The number of BCs for $\hat{U}^{(n)}$ is $k$ for all
$n$, and $\hat{U}^{(n)}$ is associated with a renormalized Planck's
constant $\hbar ^{(n)}=\hbar /k^{n}$. Thus, $\hat{U}^{(n)}$ has $k^{n}p$
eigenstates at fixed BCs. The quantum cat maps are fixed points of
either ${\cal R}$ or ${\cal R}^{2}$, so that general $\hat{U}^{(n)}$ or
$\hat{U}^{(2n)}$ represent perturbations of a given quantum cat map in
its classical limit $n\rightarrow\infty$. We then show that the
quasienergy eigenvalue problem for $\hat{U}^{(n)}$ for {\em all} $k$ BCs
is equivalent, by a unitary transformation accompanied by a scaling of
variables, to that for $\hat{U}^{(n+1)}$ at some {\em fixed} BCs. The
latter can be of four types for $n=0$ ($\hat{U}^{(0)}=\hat{U}$), but,
for $n>0$, they can be only of two types, i.e., strict {\em periodicity}
for $kp$ even and {\em antiperiodicity} for $kp$ odd. Thus, the total
(all BCs) spectrum of $\hat{U}^{(n)}$, $n=0,...,\ \overline{n}-1$
($\overline{n}>1$), coincides with a fraction $k^{1+n-\overline{n}}$ of
the spectrum of $\hat{U}^{(\overline{n})}$ for one of these two types of
BCs, and the corresponding eigenstates are related by the transformation
above. In particular, the total spectrum of a quantum cat map for $\hbar
=2\pi /p$ coincides with a fraction $k^{1-n}$ of its fixed-BCs spectrum
for $\hbar =2\pi /(k^{n}p)$, with arbitrary or even $n>0$. It is
interesting to note that the two types of BCs above are precisely those
that have been studied in detail in the literature
\cite{hb,e,k,ke,cat,n,pc,pc1,pc2}, so that several results, in
particular those of Hannay and Berry \cite{hb} for the quantum cat maps,
turn out now to be significant also for general BCs.\newline

We denote by $(u,\ v)$ the phase-space variables, $[\hat{u},\
\hat{v}]=i\hbar$, and we assume that the classical dynamics can be
reduced to a $2\pi\times 2\pi$ torus $T^{2}$, where it is described by
an Anosov map $M$. In general, a smooth torus map $M$ can be expressed
uniquely as the composition of two maps, $M=M_{A}\circ M_{1}$
\cite{kmr}. Here $M_{A}$ is a cat map, $M_{A}({\bf z)}=A\cdot {\bf z}\ \
$mod$\ 2\pi$, where ${\bf z}$ is the column vector $(u,\ v)^{T}$ and $A$
is a $2\times 2$ integer matrix with $\det (A)=1$; by ``Anosov'' we just
mean that $|$Tr$(A)|>2$, a condition generically satisfied by $A$. The
map $M_{1}$ is defined by $M_{1}({\bf z})= {\bf z}+{\bf F}({\bf z})$ mod
$2\pi$, where ${\bf F}({\bf z})$ is a $2\pi$-periodic vector function of
${\bf z}$. The QAM corresponding to $ M=M_{A}\circ M_{1}$ is the unitary
operator $\hat{U}=\hat{U}_{A}\hat{U}_{1}$ \cite{kmr}, where
$\hat{U}_{A}$ is the ``quantum cat map'', whose $u$ representation is
\cite{hb}
\begin{equation}
\langle u_{2}|\hat{U}_{A}|u_{1}\rangle_{\hbar}=\left( \frac{1}{2\pi
i\hbar A_{1,2}}\right) ^{1/2}\exp\left[ \frac{i}{2\hbar A_{1,2}}\left(
A_{1,1}u_{1}^{2}-2u_{1}u_{2}+A_{2,2}u_{2}^{2}\right) \right] \label{UA}
\end{equation}
with $\hbar =2\pi /p$ ($p$ integer), and $\hat{U}_{1}$ is the
quantization of the map $M_{1}$. We shall assume that $M_{1}$ is the map
for a Hamiltonian which is periodic in phase space with unit cell
$T^{2}$. As shown in Ref. \cite{kmr}, this is the case if and only if
$\int_{T^{2}}{\bf F}({\bf z})\,d{\bf z=0}$. Then $\hat{U}_{1}$ is the
one-step evolution operator for the Weyl quantization of this
Hamiltonian and is a periodic operator function $\hat{U}_{1}({\bf
\hat{z}};\ \hbar )$, representable by a well defined Fourier expansion.
The toral quantum states must be simultaneous eigenstates of the
commuting phase-space translations on $T^{2}$, $\hat{D}_{1}=\exp
(ip\hat{u})$ and $\hat{D}_{2}=\exp (-ip\hat{v})$; the corresponding
eigenvalues are $\exp (ipw_{1})$ and $\exp (-ipw_{2})$, where $(w_{1},\
w_{2})^{T}={\bf w}$ is the Bloch wave vector specifying the toral BCs
\cite{d1}. An eigenstate $\Psi_{{\bf w}}$ of $\hat{D}_{1}$ and
$\hat{D}_{2}$ can be an eigenstate of $\hat{U}$ only for those values of
${\bf w}$ in the Brillouin zone (BZ: $0\leq w_{1},\ w_{2}<2\pi /p$)
satisfying the equation \cite{kmr}
\begin{equation}
A\cdot {\bf w}={\bf w}+\pi {\bf y}\ \text{mod }2\pi /p, \label{cw}
\end{equation}
where ${\bf y}\equiv (A_{1,1}A_{1,2},\ A_{2,1}A_{2,2})^{T}$. We write
the general solution of Eq. (\ref{cw}) as follows:
\begin{equation}
{\bf w}=(2\pi /p)B\cdot ({\bf r}+pE^{-1}\cdot {\bf y}/2)\text{ mod }2\pi
/p, \label{sw}
\end{equation}
where $B=(A-I)^{-1}E$, $I$ is the identity matrix, $E$ is an arbitrary
integer matrix with $\det (E)=\pm 1$, and ${\bf r}$ is an integer vector
labeling the solutions. There are precisely $k=|\det
(B^{-1})|=|2-$Tr$(A)|$ distinct vectors (\ref{sw}), as the number of
fixed points of $M_{A}$ \cite{ke}, forming a {\em lattice} in the BZ. We
denote by ${\cal S}$ the space of states $\Psi_{{\bf w}}$ for all these
$k$ values of ${\bf w}$. The subspace ${\cal S}_{{\bf w}}$ of ${\cal S}$
with a fixed value of ${\bf w}$ is $p$-dimensional, i.e., it is spanned
by a basis of $p$ independent states \cite{d1,d3}, whose general
expression in the $u$ representation is \cite{note} 
\begin{equation}
\Psi_{b,{\bf w}}(u)=\sum_{m=0}^{p-1}\phi_{b}(m;\ {\bf w})
\sum_{l=-\infty}^{\infty} e^{ilpw_{2}}\delta (u-w_{1}-2\pi m/p-2\pi l),
\label{qes}
\end{equation}
where $b=1,...,\ p$. Such a basis is formed, naturally, by the $p$
eigenstates of $\hat{U}$ at fixed ${\bf w}$.\newline

We now introduce the torus $T_{B}^{2}$, defined by the vectors ${\bf
R}_{j}=2\pi k(B_{1,j},\ B_{2,j})^{T}$, $j=1,\ 2$;  $kB$ has integer
entries and $T_{B}^{2}$ contains precisely $k$ tori $T^{2}$. Since
$B^{-1}AB=E^{-1}AE$ is an integer matrix, the superlattice with unit
cell $T_{B}^{2}$ is invariant under $A$, so that the map $M$ modulo
$T_{B}^{2}$, denoted by $M^{(B)}=M_{A}^{(B)}\circ M_{1}^{(B)}$, is well
defined. To continue, we shall first work out in detail the case of
Tr$(A)<-2$, choosing $E=I$, so that $[A,\ B]=0$ and $\det (B)>0$. We
shall then specify the changes to be made in the case of Tr$(A)>2$. Let
us perform the linear transformation of variables
\begin{equation}
{\bf z}=kB\cdot {\bf z}^{\prime}=\sqrt{k}C\cdot {\bf z}^{\prime},
\label{rt}
\end{equation}
where $C=\sqrt{k}B$ and ${\bf z}^{\prime}=(u^{\prime},\
v^{\prime})^{T}$. Eq. (\ref{rt}) is the combination of a linear
canonical transformation [$\det (C)=+1$] with a scaling by a factor
$\sqrt{k}$. Using $[A,\ B]=0$, it is easy to check that the map
$M^{(B)}$ above is transformed by (\ref{rt}) into the map
$M^{\prime}=M_{A}^{\prime}\circ M_{1}^{\prime}$ on $T^{2}$ in the ${\bf
z}^{\prime}$ variables, where $M_{A}^{\prime}$ is the ordinary cat map,
$M_{A}^{\prime}({\bf z}^{\prime})=A\cdot {\bf z}^{\prime}$ mod $2\pi$,
and $M_{1}^{\prime}({\bf z}^{\prime})=(kB)^{-1}\cdot M_{1}^{(B)}({\bf
z}=kB\cdot {\bf z}^{\prime})$. The renormalization transformation ${\cal
R}_{c}$ in the classical case is then defined by $M^{\prime}({\bf
z}^{\prime})={\cal R}_{c}[M({\bf z})]$. Clearly, the cat maps
($M=M_{A}$) are fixed points of ${\cal R}_{c}$, i.e., $M^{\prime}({\bf
z}^{\prime}={\bf z})=M({\bf z})$.\newline

The quantum version of (\ref{rt}) implies that $[\hat{u}^{\prime},\
\hat{v}^{\prime}]=i\hbar ^{\prime}$, where $\hbar ^{\prime}=\hbar
/k=2\pi /p^{\prime}$, $p^{\prime}\equiv kp$. The quantization
$\hat{U}^{\prime}$ of $M^{\prime}({\bf z}^{\prime})$ is simply $\hat{U}$
expressed in terms of $({\bf \hat{z}}^{\prime},\ \hbar ^{\prime})$ and
acting on the space of the simultaneous eigenstates of the phase-space
translations on $T^{2}$ in the ${\bf z}^{\prime}$ variables,
$\hat{D}_{1}^{\prime}=\exp (ip^{\prime}\hat{u}^{\prime})$ and
$\hat{D}_{2}^{\prime}=\exp (-ip^{\prime}\hat{v}^{\prime})$. It is easy
to show that
\begin{equation}
\hat{D}_{j+1}^{\prime}=\hat{D}({\bf R}_{j})=(-1)^{pk^{2}B_{1,j}B_{2,j}}
\hat{D}_{1}^{kB_{2,j}}\hat{D}_{2}^{kB_{1,j}} \label{WH}
\end{equation}
($j=1,\ 2$, $\hat{D}_{3}^{\prime}\equiv\hat{D}_{1}^{\prime}$), where
$\hat{D}({\bf R}_{j})$ are precisely the Weyl-Heisenberg phase-space
translations on $T_{B}^{2}$. By expressing
$\hat{U}=\hat{U}_{A}\hat{U}_{1}$ in terms of $({\bf \hat{z}}^{\prime},\
\hbar ^{\prime})$, using also the theory of linear quantum canonical
transformations \cite{mq}, we obtain the expected result
$\hat{U}^{\prime}=\hat{U}_{A}^{\prime}\hat{U}_{1}^{\prime}$. Here the
$u^{\prime}$ representation of $\hat{U}_{A}^{\prime}$ is given by
(\ref{UA}) with $u$ and $\hbar$ replaced by $u^{\prime}$ and $\hbar
^{\prime}$, respectively, and $\hat{U}_{1}^{\prime}$ is the operator
function $\hat{U}_{1}^{\prime}({\bf \hat{z}}^{\prime};\ \hbar
^{\prime})=\hat{U}_{1}({\bf \hat{z}}=kB\cdot {\bf \hat{z}}^{\prime};\
\hbar =k\hbar ^{\prime})$ [the function $\hat{U}_{1}({\bf \hat{z}};\
\hbar )$ was defined above]. The renormalization transformation ${\cal
R}$ is then defined by $\hat{U}^{\prime}={\cal R}(\hat{U})$. By
iterating ${\cal R}$, one obtains a sequence of QAMs
$\hat{U}^{(n)}={\cal R}^{n}(\hat{U})$ on $T^{2}$, associated with the
Planck's constants $\hbar ^{(n)}=2\pi /p^{(n)}$, $p^{(n)}\equiv k^{n}p$.
The quantum cat maps ($\hat{U}=\hat{U}_{A}$) are fixed points of ${\cal
R}$, i.e., $\langle u_{2}^{\prime}=u_{2}|\hat{U}
^{\prime}|u_{1}^{\prime}=u_{1}\rangle_{\hbar ^{\prime}=\hbar}=\langle
u_{2}|\hat{U}|u_{1}\rangle_{\hbar}$. General
$\hat{U}^{(n)}=\hat{U}_{A}^{(n)}\hat{U}_{1}^{(n)}$ represent
perturbations of the quantum cat map $\hat{U}_{A}$ in its classical
limit $\hbar ^{(n)}\rightarrow 0$ ($n\rightarrow\infty$). The
perturbation $\hat{U}_{1}^{(n)}({\bf \hat{z}} ^{(n)};\ \hbar ^{(n)})$ is
periodic in ${\bf \hat{z}}^{(n)}$ with a unit cell $(kB)^{-n}\cdot
T^{2}$, which is $k^{n}$ times smaller than $T^{2}$.\newline

The renormalized Bloch wave vector ${\bf w}^{(n)}$ assumes again $k$
values given by Eq. (\ref{sw}) with $p$ replaced by $p^{(n)}$. Consider
first $n=1$. The space of states $\Psi_{{\bf w}^{\prime}}^{\prime}$ for
all the $k$ values of ${\bf w}^{\prime}$ will be denoted by ${\cal
S}^{\prime}$. We now show that the original space ${\cal S}$ coincides
with the subspace ${\cal S}_{{\bf w}_{0}^{\prime}}^{\prime}$ of ${\cal
S}^{\prime}$ associated with a particular allowed value ${\bf
w}_{0}^{\prime}$. Thus, ${\cal S}^{\prime}$ is $k$ times larger than
${\cal S}$. To show this, let us apply $\hat{D}_{j}^{\prime}$, $j=1,\
2$, on a state $\Psi_{{\bf w}}$ of ${\cal S}$. Using (\ref{sw}),
(\ref{WH}), and the fact that $\hat{D}_{j}\Psi_{{\bf w}}=\exp
[i(-1)^{j+1}pw_{j}]\Psi_{{\bf w}}$, $j=1,\ 2$, we obtain
\begin{equation}
\hat{D}_{j}^{\prime}\Psi_{{\bf w}}=(-1)^{pA_{j,j+1}}\Psi_{{\bf w}}
\label{Dpw}
\end{equation}
($A_{2,3}\equiv A_{2,1}$). Rel. (\ref{Dpw}) means that {\em all}
$\Psi_{{\bf w}}$ in ${\cal S}$ are eigenstates of
$\hat{D}_{j}^{\prime}$, $j=1,\ 2$, associated with the {\em same}
renormalized Bloch wave vector ${\bf w}_{0}^{\prime}$. The latter can
assume only four values, depending on the matrix $A$, see Table 1.
\[
\begin{tabular}{||c||c||c||}
\hline\hline & $k$ even & $k$ odd \\ \hline\hline $p$ even & ${\bf
w}_{0}^{\prime}={\bf 0}$ & ${\bf w}_{0}^{\prime}={\bf 0}$ \\
\hline\hline $p$ odd & ${\bf w}_{0}^{\prime}=\left(
\frac{A_{1,2}\pi}{p^{\prime}},\ \frac{A_{2,1}\pi }{p^{\prime}}\right)
^{T}$ mod $\frac{2\pi}{p^{\prime}}$ & ${\bf w}_{0}^{\prime}=\left(
\frac{\pi }{p^{\prime}},\ \frac{\pi}{p^{\prime}}\right) ^{T}$ \\
\hline\hline
\end{tabular}
\]
\[
\text{Table 1.} 
\]
It is easy to show that ${\bf w}_{0}^{\prime}$ is indeed an allowed
value of ${\bf w}^{\prime}$ in all four cases. Thus, ${\cal S}_{{\bf
w}_{0}^{\prime}}^{\prime}$ includes ${\cal S}$, but since both ${\cal
S}_{{\bf w}_{0}^{\prime}}^{\prime}$ and ${\cal S}$ are $kp$-dimensional,
they coincide. This completes the proof. Now, by the definition above of
$\hat{U}^{\prime}$, the restriction of $\hat{U}^{\prime}$ to ${\cal
S}_{{\bf w}_{0}^{\prime}}^{\prime}={\cal S}$ is just $\hat{U}$. The $kp$
eigenstates of $\hat{U}$ for all $k$ BCs are then precisely the
$p^{\prime}$ eigenstates of $\hat{U}^{\prime}$ associated with the value
of ${\bf w}_{0}^{\prime}$ in Table 1. When referred to
$\hat{U}^{\prime}$, however, these eigenstates should be expressed in a
representation based on the operator ${\bf \hat{z}}^{\prime}$. If the
$kp$ eigenstates of $\hat{U}$ are $\Psi_{b,{\bf w}}(u)$ in the $u$
representation, see (\ref{qes}), their $u^{\prime}$ representation will
be obtained by applying to $\Psi_{b,{\bf w}}(u)$ the unitary
transformation corresponding to a linear canonical transformation
\cite{mq} with matrix $C$, after scaling $u^{\prime}$ by a factor
$\sqrt{k}$. The eigenstates of $\hat{U}^{\prime}$ for ${\bf
w}^{\prime}={\bf w}_{0}^{\prime}$ are thus given by 
\begin{equation}
\Psi_{b^{\prime},{\bf w}_{0}^{\prime}}^{\prime}(u^{\prime})=\left(
\frac{p}{4\pi ^{2}B_{1,2}}\right) ^{1/2}\int\limits_{-\infty}^{\infty}
du\exp\left[ \frac{-ip}{4\pi B_{1,2}}\left(
kB_{1,1}u^{\prime}{}^{2}-2u^{\prime}u+B_{2,2}u^{2}\right) \right]
\Psi_{b,{\bf w}}(u), \label{qep}
\end{equation}
where $b^{\prime}=b^{\prime}(b,\ {\bf w})$ takes precisely all its
$p^{\prime}$ values when $b$ and ${\bf w}$ take all their $p$ and $k$
values, respectively; conversely, $\Psi_{b,{\bf w}}(u)$ can be fully
reproduced from $\Psi_{b^{\prime},{\bf
w}_{0}^{\prime}}^{\prime}(u^{\prime})$ by inverting Rel. (\ref{qep}) and
determining ${\bf w}$ by applying $\hat{D}_{1}$ and $\hat{D}_{2}$ on
$\Psi_{b,{\bf w}}(u)$. If the quasienergies of $\hat{U}$ are
$\omega_{b}({\bf w})$, those of $\hat{U}^{\prime}$ for ${\bf
w}^{\prime}={\bf w}_{0}^{\prime}$ are $\omega_{b^{\prime}(b,{\bf
w})}^{\prime}({\bf w}_{0}^{\prime})=\omega_{b}({\bf w})$. The latter
relation and Rel. (\ref{qep}) show the equivalence between the
quasienergy eigenvalue problem for $\hat{U}$ for all $k$ BCs and that
for $\hat{U}^{\prime}$ at the fixed BCs given by ${\bf w}^{\prime}={\bf
w}_{0}^{\prime}$.\newline

The generalization of these results to $n>1$ is straightforward. The
fixed BCs for $\hat{U}^{(n)}$ are determined from Table 1 with $p$,
$p^{\prime}$, and ${\bf w}_{0}^{\prime}$ replaced by $p^{(n-1)}$,
$p^{(n)}$, and ${\bf w}_{0}^{(n)}$, respectively. However, since
$p^{(n-1)}=k^{n-1}p$ is always even when $k$ is even, ${\bf
w}_{0}^{(n)}$ can take now {\em only two} values, ${\bf
w}_{0}^{(n)}={\bf 0}$ (for $kp$ even) and ${\bf w}_{0}^{(n)}=(\pi
/p^{(n)},\ \pi /p^{(n)})$ (for $kp$ odd), corresponding to strictly
periodic and antiperiodic BCs, respectively. The eigenstates of
$\hat{U}^{(n)}$ for ${\bf w}^{(n)}={\bf w}_{0}^{(n)}$ are connected with
those of $\hat{U}^{(n-1)}$ for all $k$ BCs by a relation analogous to
Rel. (\ref{qep}). The quasienergies are related by
$\omega_{b^{(n)}}^{(n)}({\bf
w}_{0}^{(n)})=\omega_{b^{(n-1)}}^{(n-1)}({\bf w}^{(n-1)})$. Thus, the
spectrum and eigenstates of $\hat{U},...,\ \hat{U}^{(n-1)}$ for all BCs
can be fully reproduced from those of $\hat{U}^{(n)}$ for ${\bf
w}^{(n)}={\bf w}_{0}^{(n)}$.\newline

The case of Tr$(A)>2$ can be treated similarly only if the integer
matrix $E$ in $B=(A-I)^{-1}E$ can be chosen so that $[A,\ E]=0$ and
$\det (E)=-1$. Then one has again $[A,\ B]=0$ and $\det (C)=+1$ in
(\ref{rt}), leading to the same results as above. If $A=K^{2l}$, where
$K$ is any integer matrix with $\det (K)=-1$ and $l$ is an integer, one
can choose $E=K$; see also the example below. In general, an integer
matrix $E$ having the properties above does not exist, and we then make
the simple choice $E_{1,1}=E_{2,2}=0$, $E_{1,2}=E_{2,1}=1$. While $[A,\
E]\neq 0$, this introduces only ``minimal'' changes which do not affect
the main general results. In all the expressions and equations involving
the $n=1$ renormalized quantities, including in Table 1, $A$ is replaced
by $A^{\prime}=E^{-1}AE$. In the next renormalization, the matrix
$B^{\prime}=(A^{\prime}-I)^{-1}E$ is used instead of $B$, starting from
Eq. (\ref{sw}), and in all the expressions and equations involving the
$n=2$ quantities $A$ is left unchanged, since
$A^{(2)}=E^{-1}A^{\prime}E=A$. In general, $A$ is replaced by
$A^{\prime}$ only for $n$ odd and $B^{\prime}$ is then used instead of
$B$ in the next, $(n+1)$th renormalization. Thus, for the quantum cat
maps, the dependence of $\hat{U}^{(n)}$ on $({\bf \hat{z}}^{(n)},\ \hbar
^{(n)})$ is the same for values of $n$ differing by an even integer and
is different otherwise due to the replacement of $A$ by $A^{\prime}$.
These maps are therefore fixed points of ${\cal R}^{2}$.\newline

As a simple example, we consider the case when the vectors (\ref{sw})
form a square lattice in the BZ, ${\bf w}=2\pi (r_{1},\
r_{2})^{T}/(gp)$, $r_{1},\ r_{2}=0,...,\ g-1$, $g$ integer. This will be
the case only if $p{\bf y}/2$ in (\ref{sw}) is an integer vector and the
matrix $E$ can be chosen so that $B=(A-I)^{-1}E=I/g$. This implies that
$A=I+gE$. It is easy to show that the latter relation is satisfied if
and only if $A=sE^{2}$ with Tr$(E)=sg$, where $s=\pm 1$ is the sign of
Tr$(A)$ [or of $-\det (E)=s$]. Using these conditions, one can easily
find matrices $A$ for any $g$. Note that in this case $\det
(B)=g^{-2}>0$, independently of $s$, so that the quantum cat maps are
always fixed points of ${\cal R}$. The transformation (\ref{rt}) is
simply ${\bf z}=g{\bf z}^{\prime}$, and the eigenstates
$\Psi_{b^{\prime},{\bf w}_{0}^{\prime}}^{\prime}(u^{\prime})$ of
$\hat{U}^{\prime}$ can be easily determined, without using (\ref{qep}),
by just substituting $u=gu^{\prime}$ in (\ref{qes}). After rearranging
terms, we find that $\Psi_{b^{\prime},{\bf
w}_{0}^{\prime}}^{\prime}(u^{\prime})$ is given by the expression in Eq.
(\ref{qes}) with all the quantities replaced by their primed
counterparts and ${\bf w}^{\prime}={\bf w}_{0}^{\prime}={\bf 0}$ (in
fact, since $p{\bf y}/2$ is an integer vector, $kp$ must be even,
implying that ${\bf w}_{0}^{(n)}={\bf 0}$ for all $n$). For given $b$
and ${\bf w}$, an expansion coefficient $\phi_{b^{\prime}(b,{\bf
w})}(m^{\prime};\ {\bf w}_{0}^{\prime})$, $m^{\prime}=0,...,\
p^{\prime}-1$, is nonzero only if there exists an integer pair $(m,\
l)$, $m=0,...,\ p-1$, $l=0,...,\ g-1$, solving the Diophantine equation
$pgl+gm+r_{1}=m^{\prime}$. The solution is then unique and
$\phi_{b^{\prime}}(m^{\prime};\ {\bf w}_{0}^{\prime})=\exp (2\pi
ilr_{2}/g)\phi_{b}(m;\ {\bf w})$, associated with a ``sparse''
expansion. In particular, for $p=1$, i.e., $\hbar ^{\prime}=2\pi
/g^{2}$, we can choose $\phi_{b}(m;\ {\bf w})=1$, and the only nonzero
coefficients are $\phi_{b^{\prime}}(m^{\prime};\ {\bf
w}_{0}^{\prime})=\exp (2\pi ilr_{2}/g)$ with $m^{\prime}=gl+r_{1}$. This
result can be obtained directly by applying the methods in Ref.
\cite{k}, where $2\pi /\hbar =$ square integer was assumed, to the case
of $A=I+gE$.\newline

In conclusion, the results in this paper provide a first understanding
of basic spectral properties of QAMs for general BCs on a torus. The
QAMs may be viewed as {\em generic} quantum torus maps, with the only
restriction $|$Tr$(A)|>2$ on their associated matrix $A$.  For given
$\hbar =2\pi /p$, this matrix determines {\em completely} the general
BCs, whose number, $k$, is always {\em finite}. This is the basis for
the renormalization scheme introduced in this paper. Thus, this scheme
cannot be applied to the special torus maps with Tr$(A)=2$ (in
particular, the periodic maps with $A=I$, such as the kicked Harper map
\cite{l,d1,d2,d3}), since $k=\infty$ for them. There is a useful freedom
in the definition of the renormalization transformation ${\cal R}$, due
to the arbitrariness in the choice of the integer matrix $E$ in
(\ref{sw}), with $\det (E)=-$sgn$[$Tr$(A)]$. For Tr$(A)<-2$ and for at
least a large class of matrices $A$ with Tr$(A)>2$, $E$ can be chosen so
that $[A,\ E]=0$. Then, the iterates $\hat{U}^{(n)}$ of a QAM $\hat{U}$
under ${\cal R}$ are perturbations of the {\em same} quantum cat map
$\hat{U}_{A}$ with a Planck's constant $\hbar ^{(n)}=\hbar /k^{n}$
tending to the classical limit as $n\rightarrow\infty$. If there exists
no $E$ satisfying $[A,\ E]=0$ for Tr$(A)>2$, $E$ can be chosen so that
$\hat{U}_{A}$ is a fixed point of ${\cal R}^{2}$. The main general
result, however, is the same for all $A$: The spectrum and eigenstates
of $\hat{U},...,\ \hat{U}^{(n-1)}$ for {\em all} BCs can be fully
reproduced from those of $\hat{U}^{(n)}$ at some special, {\em fixed}
BCs. The general BCs are then practically ``eliminated'' by ${\cal R}$
up to an arbitrarily high order $n$. For $n>1$, the fixed BCs can be
only of two types: strict periodicity for $kp$ even and antiperiodicity
for $kp$ odd. Thus, several previous results
\cite{hb,e,k,ke,cat,n,pc,pc1,pc2} for these special BCs turn out now to
be significant also for general BCs. In addition, some results that can
be obtained by previous methods at fixed BCs, such as those in Ref.
\cite{k}, may now be understood better and derived in a simpler way in
the general BCs framework, as in the example above. Finally, the
renormalization scheme may be used to study the sensitivity of the
eigenstates to variations in the BCs and to derive new exact results
concerning, for example, the influence of the general BCs on the
spectral degeneracies in the quantum cat maps and on the removal of
these degeneracies by small perturbations.\newline

{\bf Acknowledgments}\newline

The author would like to thank J.P. Keating and Z. Rudnick for
discussions. This work was partially supported by the Israel Science
Foundation administered by the Israel Academy of Sciences and
Humanities.

\end{document}